\documentstyle[multicol,prl,aps,psfig,amsfonts]{revtex}
\begin{document}
\draft
\title{Critical behaviour at the metal-insulator transition in 3-dimensional 
disordered systems in a strong magnetic field.}
\author{E. Hofstetter}
\address{Blackett Laboratory, Imperial College, London SW7 2BZ, UK}
\date{\today}
\maketitle
\begin{abstract}
The critical behaviour of 3-dimensional disordered systems in a strong magnetic 
field is investigated by analysing the spectral fluctuations of the energy 
spectrum. The level spacing distribution $P(s)$ as well the Dyson-Mehta 
statistics $\Delta_{3}(L)$ are considered. Except for the small $s$ behaviour
of $P(s)$, which depends on the presence or absence of a magnetic field but not 
on its strength, the other quantities, large $s$ behaviour of $P(s)$ and 
$\Delta_{3}(L)$, turn out to be independent of its presence or absence and 
therefore of the time reversal symmetry. This suggests that the metal-insulator
transition, which is defined by the symmetry of the system {\it at} the critical
point, is independent of the presence or absence of the time reversal symmetry.
\end{abstract}
\vspace{1cm}
\pacs{PACS numbers: 71.30.+h, 05.45.+b, 64.60.Cn }

\begin{multicols}{2}
It is now well know that 3-dimensional disordered systems exhibit a 
metal-insulator transition (MIT) as a function of the disorder. Although a lot
of work has already been devoted to this problem \cite{Ang1} many points still 
remain unclear. One of them is the influence of the magnetic field at the MIT.
In 2-dimensional systems the magnetic field has a dramatic effect. Whereas no 
MIT is expected without a magnetic field it turns out that systems in a magnetic 
field exhibit one. Delocalised states appear at the centre of the Landau 
bands and quantities like the critical disorder or the critical exponent
can be computed \cite{Bodo}. In 3-dimensional systems it was assumed, although 
there is no such dramatic effect, that the critical behaviour at the MIT can be 
cast into three different universality classes according to some general symmetries
of the system: orthogonal (with time reversal symmetry), unitary (without time 
reversal symmetry, e.g. with a magnetic field) and symplectic (with spin 
orbit-coupling) \cite{Ang1}. One then expects different critical exponents
related to the MIT for the three different universality classes. Surprisingly,
in spite of the change of universality class, the same value of the critical
exponent has been found, numerically, with and without a magnetic field 
\cite{Kra1,Kra2}.
A convenient way to study this problem is to resort to random matrix theory
(RMT) and energy level statistics (ELS). It has been shown \cite{Eh1,Shkl},
in the case without magnetic field, that beside the two expected statistics,
namely the Gaussian orthogonal ensemble (GOE) for the metallic regime and the 
Poisson ensemble (PE) for the insulating regime, a third statistic, called the
critical ensemble (CE), occurs only exactly at the critical point. It then 
seems natural that the universality class of the system should be determined 
by the symmetry of the system {\it at} the critical point with the results
quoted above \cite{Kra1,Kra2} being interpreted as the sign of the independence
of the universality class, {\it at} the critical point, to teh breaking of  
time reversal symmetry. This has been checked in the case of a system subject 
to an Aharonov-Bohm (AB)-flux \cite{Eh2}. Despite the fact there is no magnetic
field inside the system the AB-flux breaks the time reversal symmetry and one
observes a transition this time from the Gaussian unitary ensemble (GUE) to
the PE. But the CE, the critical exponents, does not depend on the time reversal symmetry and so 
\par
A reservation which has to be considered with this kind of system is that a 
change of the 
vector potential is, by means of a gauge transformation, equivalent to a change 
of the boundary conditions and therefore should have no influence in the localised
regime. Despite this it must be stressed that, when using periodic 
boundary conditions, the vector potential cannot be gauged away and the system 
is not equivalent to a system invariant under orthogonal transformation. 
Nevertheless, in this paper we would like to address the problem now with a 
magnetic field inside the system and to show that the conclusions derived from 
the case with the AB-flux still remain valid.
\par
In order to investigate the MIT with a magnetic field we consider the usual
Anderson Hamiltonian,
\begin{equation}
H=\sum_{{\underline i}} \varepsilon_{{\underline i}} |{\underline i}\rangle
\langle {\underline i}|+\sum_{{\underline i},{\underline i'}}
V_{{\underline i},{\underline i'}} |{\underline i}\rangle
\langle {\underline i'}| \;,
\end{equation}
where the sites {\it {\underline i}=lmn} are distributed regularly in 3D space, 
e.g. on a simple cubic lattice. Only interactions with the nearest neighbours 
are considered. The site energy $\varepsilon_{lmn}$ is described by a stochastic 
variable. In the present investigation we use a box distribution with variance
$\sigma^{2}=W^{2}/12$. $W$ represents the disorder and is the critical parameter.
Introducing a homogeneous magnetic field ${\bf B}=B(1,1,1)$ leads to complex 
hopping elements (Peierls factors)\cite{Lutt}. Using the following vector 
potential ${\bf A}=B(z,x,y)$ and imposing $|V_{lmn,l'm'n'}|=1$, which defines 
the energy scale, yields for $V_{lmn,l'm'n'}$:
\begin{equation}
\left\{
\begin{array}{ll}
\exp(\pm i \frac{2\pi a^{2}B m}{h}) & {\rm if}\; l'=l\pm 1,m'=m,n'=n \\
\exp(\pm i \frac{2\pi a^{2}B n}{h}) & {\rm if}\; l'=l,m'=m\pm 1,n'=n \\
\exp(\pm i \frac{2\pi a^{2}B l}{h}) & {\rm if}\; l'=l,m'=m,n'=n\pm 1 \;,
\end{array}
\right.
\end{equation}
where $a$ is the step of the lattice. $\alpha=\frac{2\pi a^{2}B}{h}$ is the
the number of flux quanta $h/e$ per plaquette. The boundary conditions are
taken to be periodic. 
\par
Based on this Hamiltonian, the MIT in the presence of a magnetic field will be
studied by the ELS method, i.e. via the fluctuations of the energy spectrum
\cite{Eh1}. Before giving the results we shortly review the ELS method. Starting
from Eq.(1) the energy spectrum was computed by means of the Lanczos algorithm 
(which is suited to diagonalise such very sparse secular matrices) for systems 
of size $M \times M \times M$ with $M=13,\;17, \;{\rm and}\; 21$, disorder $W$ 
ranging from 3 to 40 and phase $\alpha$ from 0.005 to 0.250. The number of 
different realizations of the random site energies $\varepsilon_{lmn}$ was chosen 
so that about $ 10^{5}$ eigenvalues were obtained for every pair of parameters 
$(M,W)$ which means between 25 and 90 realizations, for which only half of the 
spectrum around the band centre is considered so that the results do not 
deteriorate due to the strongly localised states near the band edges. After 
unfolding the spectrum obtained, the fluctuations can be appropriately 
characterised \cite{OB} by means of the spacing distribution $P(s)$ and the 
Dyson-Mehta statistics $\Delta_{3}(L)$. $P(s)$ measures the level repulsion, it 
is normalised, as is its first moment, because the spectrum is unfolded. 
$\Delta_{3}(L)$ measures the spectral rigidity. 
%
%
\begin{figure}
\centerline{
\psfig{file=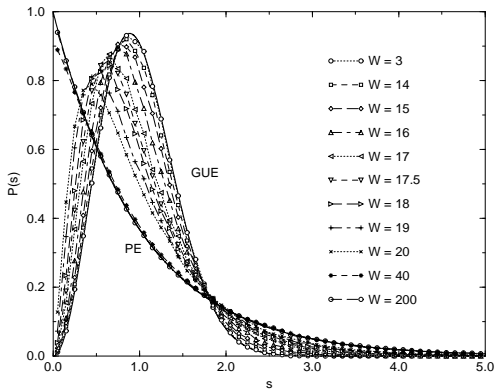,width=3.0in}
}
\vspace{0.08in}
\setlength{\columnwidth}{3.2in}
\centerline{\caption{Spacing distribution, $P(s)$, for $M=21$ and different 
disorders $W$. The full curves give the asymptotic regimes namely, GUE for the 
diffusive regime and PE for the localised regime. We see the transition between 
GUE and PE as a function of the disorder $W$.
\label{1}
}}
\end{figure}
In Fig. \ref{1} the results for the spacing distribution $P(s)$ are reported.
One finds, as expected, the GUE and the PE regimes for small and large disorder 
respectively as well as the transition between them as a function of $W$. But 
less expected is the fact that, in contrast to the case with time reversal 
symmetry \cite{Eh1,Shkl}, there is no longer one point where all the curves intersect
but rather one which seems to appear only for the curves in the metallic regime.
\par
\begin{figure}
\centerline{
\psfig{file=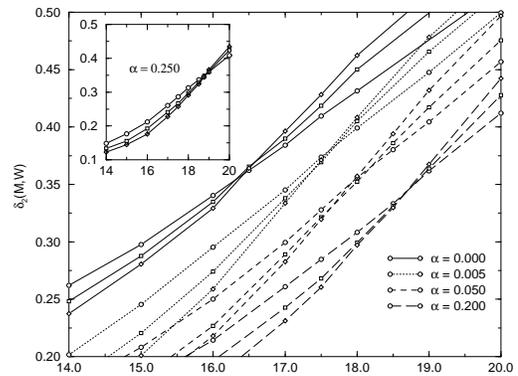,width=3.0in}
}
\vspace{0.08in}
\setlength{\columnwidth}{3.2in}
\centerline{\caption{$\delta$ as a function of $M$ and $W$. $\circ$ : $M = 13$, $\Box$ :
$M = 17$ and $\diamond$ : $M = 21$. The critical disorder is given by the point
for which $\delta$ is independent of $M$. $W_{c}$ ranges from $16.5$ without a
magnetic field up to $18.75$ where $W_{c}$ seems to become independent of the
strength of the magnetic field.
\label{2}
}}
\end{figure}
\begin{figure}
\centerline{
\psfig{file=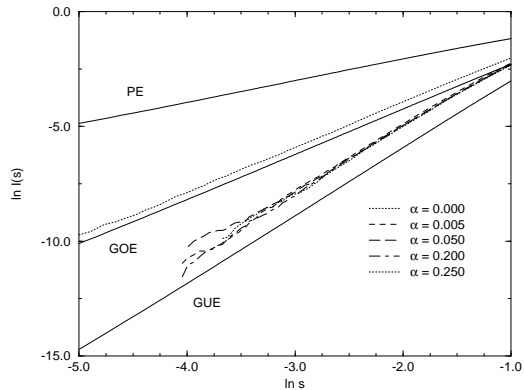,width=3.0in}
}
\vspace{0.08in}
\setlength{\columnwidth}{3.2in}
\centerline{\caption{ln-ln plot of the cumulative level-spacing distribution $I(s)$ for 
small $s$. The curves depends on the absence or presence of a magnetic field 
but are independent of its strength. In both cases the $s$ behaviour is the same
as in the metallic regime but the constants in front of $s$ and $s^{2}$ are
higher, showing a decrease of the level repulsion.
\label{3}
}}
\end{figure}

The next step is now to find where the MIT takes place. For this one uses the
fact that the quantities we are considering here, $P(s)$ and $\Delta_{3}(L)$, 
depend on the system size except {\it at} the critical point where they are 
scale invariant. This is directly related to the fact that the MIT is a second 
order transition and that finite-size scaling laws will apply close to the 
transition. From a numerical point view it has been shown that $\Delta_{3}(L)$
gives more accurate results. So we calculate $\delta (M,W)=\frac{1}{30} 
\int_{0}^{30} \Delta_{3}(L) dL$ as a function of $M$ and $W$, the critical 
disorder, $W_{c}$, being given by $W$ for which $\delta (M,W)$ is independent of
$M$ \cite{Eh3}. The results are shown in Fig. \ref{2}.
%
%
In contrast to the
results obtained with an AB-flux \cite{Eh2}, $W_{c}$ depends on the vector 
potential. This is due to the fact that now we have a magnetic field inside
the system which breaks the weak localisation phenomena \cite{Ang1}. It is then
more difficult to localise the states which means an upwards shift of $W_{c}$.
$W_{c}$ ranges now from $16.5$ without magnetic field up to $18.75$ where 
$W_{c}$ seems to become independent of the strength of the magnetic field. A 
similar behaviour has already been observed by Henneke {\it et al} \cite{Kra2}.
With $W_{c}$ it is now possible to study $P(s)$ and $\Delta_{3}(L)$ at the
critical point. The first feature we would like to consider is the small $s$
behaviour of $P(s)$. In fact for numerical reasons what we calculate is the
cumulative level-spacing distribution $I(s)=\int_{0}^{s} P(s') ds'$ which allows
a better study of the small $s$ behaviour of $P(s)$. The results are plotted
in Fig. \ref{3}. As we can see the curves depend on the absence or presence of 
a magnetic field but are independent of its strength. 
%
%
One finds, for small
$s$, $P(s)\propto a s^{2}$ with magnetic field and $P(s)\propto b s$ without. The 
behaviour is the same as in the metallic regime except for the pre-factors $a$ 
and $b$ which are now both higher showing a decrease of the level repulsion. This 
comes from the nature of the critical states which are multi-fractal \cite{Mich}.
The overlap of the wave functions between the different sites, although still 
present, is now much weaker than in the metallic regime which decreases the level
repulsion between the states. 
%
%
\begin{figure}
\centerline{
\psfig{file=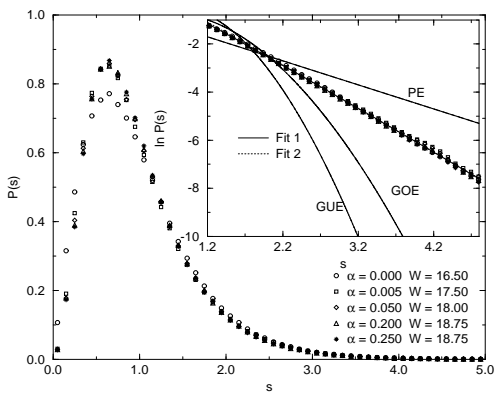,width=3.0in}
}
\vspace{0.08in}
\setlength{\columnwidth}{3.2in}
\centerline{\caption{Large $s$ behaviour of $P(s)$. The curves are no longer only 
independent of the strength of the magnetic field as for small $s$ but are
independent of the presence or absence of the magnetic field. In the inset
$\exp-As$, with $A\simeq 1.8$, (Fit 2) and $\exp-(A's+Cs^{1.25})$, with 
$A'\simeq 1.37$ and $C\simeq 0.26$, (Fit 1) have been fitted to the tail of 
$P(s)$.
\label{4}
}}
\end{figure}
In Fig. \ref{4} the other limit, namely the large $s$ behaviour of $P(s)$, is
reported. The curves are no longer only independent of the strength of the magnetic
field, as for small $s$, but are even independent of the presence or absence of 
the magnetic field. The question of the shape, for large $s$, of $P(s)$ is 
rather controversial. It was first claimed \cite{Shkl} that $P(s)$ should be 
$\propto\exp -As$. Then following some new analytical calculations at the 
critical point Aronov {\it et al.} \cite{Aro} proposed that $P(s)=B 
s^{\beta}\exp(-A\beta s^{2-\gamma})$ with $\gamma = 1-1/\nu d$ where $\nu$ is 
the critical exponent of the correlation length $\xi_{\infty}$. It has to be noted 
that some doubts have been recently raised concerning the validity of these
results \cite{Chen}. Numerically the situation is 
not really clearer. When one tries to fit a curve like $P(s)=Bs\exp(-A s^{\eta})$
for the case without magnetic field one finds an $\eta$ between 1.20 and 1.30 
\cite{Eh1,Eh4} in a good agreement with the results obtains by Aronov 
{\it et al.} \cite{Aro}. But when one looks just at the tail of $P(s)$ one can 
see some discrepancies between the fit and the numerical results. Following
some other numerical calculations it was then claimed \cite{Isa} that $P(s)$ 
was indeed $\propto\exp -As$ as first suggested \cite{Shkl}. But the fit was
done only for the large $s$ and not on the whole curve as we did previously
\cite{Eh1,Eh4}. In the inset of Fig. \ref{4} we show that, when considering only
the tail of $P(s)$, $\exp-As$ with $A\simeq 1.8$ (Fit 2) gives indeed a good 
result but so does $\exp-(A's+Cs^{1.25})$ (Fit 1). It is therefore very difficult 
to draw a definitive conclusion about the problem just based on the fit of the 
tail of $P(s)$. In the case with a magnetic field it turned out to be very 
difficult to fit a whole curve using the result of Aronov {\it et al.} which
could be related, somehow, to the absence of a fixed point in $P(s)$ as noticed
above. It clearly shows that more work needs to be done in order to understand 
properly the shape of $P(s)$ at the critical point.
%
%
\begin{figure}
\centerline{
\psfig{file=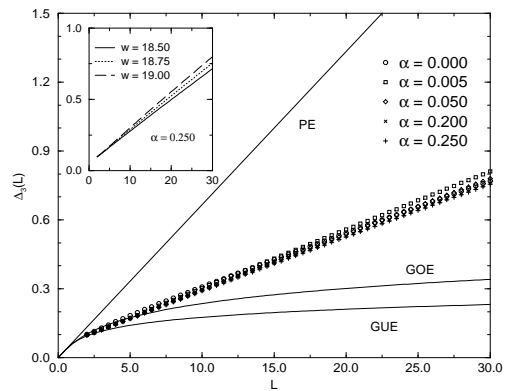,width=3.0in}
}
\vspace{0.08in}
\setlength{\columnwidth}{3.2in}
\centerline{\caption{Dyson-Mehta statistics ($\Delta_{3}(L)$) at the critical 
point for various magnetic fields. As for the large $s$ behaviour of $P(s)$, 
$\Delta_{3}(L)$ is not only independent of the strength of the magnetic field 
but of its presence or absence too.
\label{5}
}}
\end{figure}
Finally we compare $\Delta_{3}(L)$ for the different magnetic fields at the
critical point in Fig. \ref{5}. As for the large $s$ behaviour of $P(s)$, 
$\Delta_{3}(L)$ is not only independent of the strength of the magnetic field but 
seems to be independent of its presence or absence too. The shape of 
$\Delta_{3}(L)$ contains a term linear in $L$ as well as a non-linear term
$L^{\omega}$, with $\omega < 1$ and has already been studied in a previous work
\cite{Eh2}. In the inset of Fig.
\ref{5} we show how $\Delta_{3}(L)$ varies, when we slightly change $W$, to 
give an idea about the accuracy of the curves keeping in mind that the 
uncertainty for $W_{c}$ is $\pm 0.25$.
These results suggest that, although the magnetic field brings some modifications
at the critical point, as for the critical disorder $W_{c}$ or the small $s$
behaviour of $P(s)$, the symmetry of the system is indeed independent of the magnetic 
field and then of the time reversal symmetry. What happens is, when the system 
reaches the critical point, for increasing disorder, the ${\rm O}(N)$ or 
${\rm U}(N)$ symmetry is spontaneously broken. The reason for this break is 
obvious when one considers the localised regime. If the matrix would be 
${\rm O}(N)$  or ${\rm U}(N)$ invariant it would then be possible to construct 
delocalised states by linear combinations of the localised ones which is clearly 
not allowed. In RMT this phenomena has led to the concept of a preferential basis
\cite{Sha}. The distribution of the eigenvalues and the eigenvectors, which are
independent for the GOE or GUE, start to couple when the disorder is increased
and eventually break the ${\rm O}(N)$  or ${\rm U}(N)$ invariance. This breaking 
of the symmetry appears in the large $s$ behaviour of $P(s)$ explaining why
one can have a symmetry at the critical point which is independent of the 
magnetic field whereas the small $s$ behaviour of $P(s)$ is not. It has to be 
noted that the breaking of the ${\rm O}(N)$  or ${\rm U}(N)$ invariance is at 
the origin of some interesting new results related with the lifting of the sum 
rule prohibition that is observed when studying the two-level correlation 
function \cite{Can}.
\par
In summary, by means of the ELS method, we have studied the statistical 
properties of the energy spectrum of the Anderson model in a strong magnetic
field at the MIT. For this we have considered the level spacing distribution
$P(s)$ as well as the Dyson-Mehta statistics, $\Delta_{3}(L)$. Except for the 
small $s$ behaviour of $P(s)$, which depends on the presence or absence of a 
magnetic field but not on its strength, the other quantities studied, the large 
$s$ behaviour of $P(s)$ and $\Delta_{3}(L)$ turn out to be independent of its 
presence or absence 
and thus of the time reversal symmetry. This suggests that the MIT, which is 
defined by the symmetry of the system {\it at} the critical point, is defined 
by a new universality (CE) class which is independent of the presence or absence 
of time reversal symmetry.
\par
Useful discussions with A. MacKinnon are gratefully acknowledged.

\end{multicols}

\begin{thebibliography}{the ref}
\bibitem{Ang1}
B. Kramer and A. MacKinnon, Rep. Prog. Phys. {\bf 56}, 1469 (1993) and 
references therein.
\bibitem{Bodo}
B. Huckenstein, Rev. Mod. Phys. {\bf 67}, 357 (1995) and references therein.
\bibitem{Lutt}
J. M. Luttinger, Phys. Rev. B {\bf 4}, 814 (1951).
\bibitem{Kra1}
T. Ohtsuki, B. Kramer and Y. Ono, J. Phys. Soc. Jpn. {\bf 62}, 224 (1993).
\bibitem{Kra2}
M. Henneke, B. Kramer and T. Ohtsuki, Europhys. Lett. {\bf 27}, 389 (1994).
\bibitem{Pran}
D. R. Grempel, R. E. Prange and S. Fishman, Phys. Rev. A {\bf 29}, 1639 (1984).
\bibitem{Alt1}
B. L. Alt'shuler and B. I. Shklovskii, Sov. Phys. JETP {\bf 64}, 127 (1986).
\bibitem{Eh1}
E. Hofstetter and M. Schreiber, Phys. Rev. B {\bf 48}, 16979 (1993).
\bibitem{Shkl}
B. I. Shklovskii, B. Shapiro, B. R. Sears, P. Lambrianides and H. B. Shore,
Phys. Rev. B {\bf 47}, 11487 (1993).
\bibitem{Eh2}
E. Hofstetter and M. Schreiber, Phys. Rev. Lett. {\bf 73}, 3137 (1994).
\bibitem{OB}
O. Bohigias and J.-M. Giannoni, in {\it Mathematical and Computational
Methods in Nuclear Physics}, eds. J. Dehesa, J. Gomez and A. Polls,
Lecture Notes in Physics {\bf 209}, 1 (Springer, Berlin, 1984).
\bibitem{Eh3}
E. Hofstetter and M. Schreiber, Phys. Rev. B {\bf 49}, 14726 (1994).
\bibitem{Mich}
M. Schreiber, Phys. Rev. B {\bf 31}, 6146 (1985); M. Schreiber and 
H. Gru\ss bach, Phys. Rev. Lett. {\bf 67}, 607 (1991).
\bibitem{Aro}
A. G. Aronov, V. E. Kravtzov and I. V. Lerner, JETP Lett. {\bf 50}, 39
(1994).
\bibitem{Chen} Y. Chen, Preprint Imperial College.
\bibitem{Eh4}
I. Varga, E. Hofstetter, M. Schreiber and J. Pipek, Phys. Rev. B {\bf 52}, 
7783 (1995).
\bibitem{Isa} I. Zharekeshev and B. Kramer, Jpn. Appl. Phys. {\bf 34} (8B),
4361 (1995).
\bibitem{Sha}
M. Moshe, H. Neuberger and B. Shapiro, Phys. Rev. Lett. {\bf 73}, 1497 (1994);
J. -L. Pichard and B. Shapiro, J. Phys. {\bf 4}, 623 (1994).
\bibitem{Can}
A. G. Aronov adn A. D. Mirlin, Rev. B {\bf 51}, 6135 (1995);
C. M. Canali and V. E. Kravtsov, Phys. Rev. E {\bf 51}, 5185 (1995);
C. M. Canali, Phys. Rev. B {\bf 53}, 3713 (1996).
\end{thebibliography}
\end{document}